\documentclass[%
reprint,
superscriptaddress,
longbibliography,
amsmath,amssymb,
aps,
pra,
]{revtex4-2}

\usepackage{xcolor}
\usepackage{graphicx}% Include figure files
\usepackage{dcolumn}% Align table columns on decimal point
\usepackage{bm}% bold math
\usepackage{hyperref}% add hypertext capabilities
\usepackage{svg}

\newcommand{\ket}[1]{| #1 \rangle}
\newcommand{\expval}[1]{\langle #1 \rangle}
\newcommand{\Schrodinger}{Schr\"{o}dinger }

\newcommand{\addcqt}{Centre for Quantum Technologies, National University of Singapore, 3 Science Drive 2, Singapore 117543}
\newcommand{\addtuc}{School of Electrical and Computer Engineering, Technical University of Crete, Chania, Greece 73100}

\newcommand{\addangel}{AngelQ Quantum Computing, 531A Upper Cross Street, \#04-95 Hong Lim Complex, Singapore 051531}

\begin{document}

\title{Landscape approximation of low energy solutions to binary optimization problems}
% need to fit quantum into the title as well

% Quantum approximate optimization using the localization landscape

% Localization landscape of quantum bimary optimization problems

\author{Benjamin Y.L. Tan}
\email[]{b.tan@u.nus.edu}
\affiliation{\addcqt}

\author{Beng Yee Gan}
\affiliation{\addcqt}

\author{Daniel Leykam}
\email[]{daniel.leykam@gmail.com}
\affiliation{\addcqt}

\author{Dimitris G. Angelakis}
\email[]{dimitris.angelakis@gmail.com}
\affiliation{\addcqt}
\affiliation{\addtuc}
\affiliation{\addangel}

\date{\today}

\begin{abstract}
We show how the localization landscape, originally introduced to bound low energy eigenstates of disordered wave media and many-body quantum systems, can form the basis for hardware-efficient quantum algorithms for solving binary optimization problems. Many binary optimization problems can be cast as finding low-energy eigenstates of Ising Hamiltonians. First, we apply specific perturbations to the Ising Hamiltonian such that the low energy modes are bounded by the localization landscape. 
Next, we demonstrate how a variational method can be used to prepare and sample from the peaks of the localization landscape.
%Next, we propose shallow variational quantum circuits for sampling from the peaks of the localization landscape. 
Numerical simulations of problems of up to $10$ binary variables show that the localization landscape-based sampling can outperform QAOA circuits of similar depth, as measured in terms of the probability of sampling the exact ground state.
% first draft abstract, needs a lot of polishing...
\end{abstract}

\maketitle

\section{Introduction}
\label{intro}

Finding optimal solutions to Quadratic Unconstrained Binary Optimization (QUBO) problems is one proposed near term application of quantum computers \cite{bharti2022noisy}. Solving large-scale QUBO problems has importance scheduling and allocation tasks \cite{glover2019tutorial, pakhomchik2022solving, a12110224}, machine learning \cite{date2021qubo, bauckhage2019qubo, neven2008training, bapst2020pattern, matsumoto2022distance}, amongst others \cite{glover2022applications, Guan_2021, e24111685, wang2014position}. %Optimization problems in these areas are constructed in the form of a QUBO problem before applying the available heuristics to find approximate solutions to the QUBO problem and by extention, the original optimization problem of interest. 
The search for these optimal solutions is generally difficult, as QUBO problems are NP-hard \cite{fu1986application, papadimitriou2003computational, barahona1982computational}. However, in many cases obtaining approximate solutions close to the optimal can be sufficient. This is especially true within the context of industry applications where a higher quality solution, despite being sub-optimal, may still result in significant cost savings \cite{hochba1997approximation, lenstra1990approximation, williamson2011design}. 
%{\bf This paragraph needs a few references (can be review articles).}

Commonly employed techniques for solving QUBO problems using quantum computers are typically based on mapping the QUBO problem at hand to an Ising Hamiltonian, solving the problem by finding the ground state of the Ising Hamiltonian. 
Quantum algorithms to find the ground state include Quantum Annealing \cite{rajak2023quantum, santoro2006optimization, das2008colloquium}, variational problem-specific algorithms such as the Quantum Approximate Optimization Algorithm (QAOA) and its generalizations \cite{farhi2014quantum, hadfield2019quantum}, Variational Quantum Eigensolvers \cite{Peruzzo_2014, McClean_2016, 8728102}, and Quantum Assisted methods \cite{bharti2022noisy, Kyriienko_2020, Bharti_2021}.
% In general, finding approximate solutions to a QUBO problem provides little information on how other solutions with similar or better quality can be obtained. (note quantum annealing with a finite time annealing schedule will produce a superposition of low-lying eigenstates). 
% We should also discuss here practical implementation issues - QAOA / QA will converge to the exact ground state only in the limit of an infinite circuit depth. In many applications it can be useful to obtain a solution in a finite time, these methods will ideally give a mixture of low-lying states. Hyper-parameters become important - the annealing schedule / path, mixer Hamiltonian (QAOA), annealing schedule / number of QAOA steps all affect the quality of the obtained solutions. E.g. sampling frequency thresholds for QAOA~\cite{lykov2022sampling}. Thus, the ``standard'' quantum algorithms for solving QUBO also have hyper-parameters that need to be optimized.

Methods such as Quantum Annealing and QAOA have shown provable convergence to the exact ground state in the limit of infinite annealing time and circuit depth. In many applications it is necessary to obtain a solution in a finite time, in which case these methods will ideally give a mixture of low-lying states. Hyperparameters such as the annealing schedule/path, mixer Hamiltonian, number of QAOA steps, and the sampling frequency thresholds for QAOA can affect the quality of the obtained solutions~\cite{lykov2022sampling}.

Here we consider a different approach. 
Instead of an exact method that, when run on a finite-sized circuit, gives approximate solutions whose quality is difficult to predict, we consider a scheme to sample from low energy solutions with well-defined bounds, to solve the QUBO problem approximately using shallower-depth circuits. Our approach is inspired by the ``localization landscape'' used to study the Anderson localization of low energy modes of disordered systems.
% For QUBO problems with degenerate ground states, such as the MaxCut problem, a linear combination of these ground states will also minimize the energy the Hamiltonian, and sampling from such a state will yield the various solutions that correspond to a ground state.

Anderson localization is the phenomenon where eigenfunction solutions to the \Schrodinger equation with disordered potentials are confined due to wave interference~\cite{anderson1958absence}. Finding the locations where these quantum states localize typically requires solving the eigenvalue problem, as there is often seemingly little correlation between the potentials and the subregions where the peaks of these eigenfunctions can be found. Efforts into identifying these regions of localization resulted in the localization landscape (LL) function \cite{filoche2012universal}. 

The localization landscape is a function that places a tight bound on the subregions where low energy states tend to lie. 
The inverse of the landscape function serves as an effective potential that can be used to predict areas of confinement for low energy eigenstates by identifying valleys within this effective potential. 
% Since its introduction, efforts have gone into using the localization landscape to (estimate eigenvalue spectra in a bunch of works) \textcolor{cyan}{citation needed}. 
Since its introduction, efforts have gone into using the localization landscape to obtain the integrated density of states, thereby giving an estimate for the energies of the lowest eigenstates for the 1D tight binding model ~\cite{doi:10.1137/17M1156721,DAVID2021107946}. The original localization landscape function loses its accuracy when attempting to accurately identify localized regions of higher energy eigenstates, motivating the development of related landscape functions such as the $\mathcal{L}^2$ landscape~\cite{Herviou_2020}. The $\mathcal{L}^2$ landscape is able to provide a tight bound on the localization of mid-spectrum eigenstates, and can be efficiently computed with a stochastic procedure using sparse matrix methods~\cite{doi:10.7566/JPSJ.92.054707}.

Remarkably, landscape functions can be generalized beyond low-dimensional disordered systems to more general families of real symmetric matrices ($M$-matrices)~\cite{filoche2021effective}. The localization landscape theory has also been applied to many-body quantum systems ~\cite{PhysRevB.101.014201}, extending many of its well-known properties to Hamiltonians describing interacting spins, enabling the identification of regions of Hilbert space where the low-energy many-body eigenstates localize. Qualitative changes in the shape of the landscape, e.g. quantified using methods such as persistent homology, can be used as indicators of phase transitions in many-body quantum systems~\cite{hamilton2023analysis}.

In this work, we present a method of using the localization landscape to prepare a quantum state from which low energy solutions to QUBO problems can be sampled with higher probabilities. We describe how this quantum state can be prepared on a near term quantum device, and demonstrate our methods for two problem instances --- a non-degenerate randomly generated QUBO, and a degenerate MaxCut problem \cite{garey1974some, glover2019tutorial}. 

The outline of this paper is as follows: Sec.~\ref{LL} reviews the localization landscape and its application to Anderson localization and many-body localization. 
%Sec.~\ref{secQUBO} 
Sec.~\ref{samplingll} presents the mapping of QUBO problems to Ising Hamiltonians, showing how the Ising Hamiltonian can be perturbed such that its low-energy eigenstates are bounded by the localization landscape and proposing a heuristic using shallow variational circuits for sampling from this landscape suitable for noisy intermediate-scale quantum (NISQ) devices. % in Sec.~\ref{constraints}. 
Sec.~\ref{optresults} presents numerical simulations showing how the method can be used to sample low energy solutions with higher probability than shallow QAOA circuits. We analyze the effect of the two hyperparameters of the method (the energy offset and coupling strength) in Sec.~\ref{hyperparameters} before concluding with Sec.~\ref{conclusion}.
% \section{Background}
% \label{background}

% In this section, we provide an overview of Quadratic Unconstrained Binary Optimization, and the Localization Landscape.

\section{Localization Landscape}
\label{LL}

Given a disordered Hamiltonian $\hat{H}$, finding the regions where eigenstates localize typically requires solving the eigenvalue equation. 
However, Ref.~\cite{filoche2012universal} introduced a function called the \textit{localization landscape}, $u$, that is able to predict these subregions where the eigenstates of $\hat{H}$ peak at, with the requirement that its inverse is non-negative, i.e. $\hat{H}^{-1}\geq0$.
The landscape function $u$ is the solution to the following differential equation
\begin{equation}
\hat{H}u = \vec{1}
\label{Eq:LL}
\end{equation} 
where $\vec{1}$ is a vector of all $1$'s. For an eigenstate $| \phi^\beta \rangle$ of $\hat{H}$ expressed in an orthonormal basis $\{\ket{J}\}$ with energy $E^\beta$, $u$ can be expressed as \cite{Herviou_2020}: 

\begin{equation}
u_J = \sum_\beta \frac{ \langle J | \phi^\beta \rangle }{E^\beta} \sum_m  \langle m | \phi^\beta \rangle.
\label{Eq:u_j}
\end{equation} 
where $u_J$ is the $J^{\textrm{th}}$ component of $u$ and the summation is performed over all the basis states. 

Originally developed to predict areas of localization for a single particle system in a random potential with Dirichlet boundary conditions, $u$ has the useful property of being able to bound the eigenstate amplitudes according to their energies.

An effective potential, $W$, can be defined from the inverse of the landscape function $W = \frac{1}{u}$ and the regions where low energy eigenstates peak at correspond to minima in $W$, providing greater insights into the regions of localizations compared to the original potentials, which are seemingly uncorrelated to these regions.  

Ref.~\cite{PhysRevB.101.014201} extended the concept of a localization landscape to many-body systems, showing that $u$ bounds the eigenstate amplitudes of $\hat{H}$ according to:
\begin{align}
    \left| \langle J | \phi^\beta \rangle \right| & = \left| E^\beta \right| \left| \sum_m \left(\hat{H}^{-1}\right)_{Jm} \langle m |   \phi^\beta \rangle\right| \\
    & \leq \left| E^\beta \right| \| \vec{\phi}^\beta \|_\infty \sum_m \left(\hat{H}^{-1}\right)_{Jm} \label{Eq:LLineq}\\
    & = \left| E^\beta \right| \| \vec{\phi}^\beta \|_\infty u_J \label{Eq:LLbounds}
\end{align}
where $\|\vec{\phi}^\beta\|_\infty = \max_m (|\langle m | \phi^\beta \rangle|)$ is the infinity norm of $\vec{\phi}^\beta$, and the definition of $u$ in Eq.~\eqref{Eq:LL} was used to get from Eq.~\eqref{Eq:LLineq} to Eq.~\eqref{Eq:LLbounds}.
% $ = \left(\sum_i \left|\phi^\beta\ \right|^p \right)^{1/p}$$\ell_p$ norm
% \textbf{(BY: I think we can be more specific here. Just quote that it is the infinite norm. And I think $\phi^\beta$ in Eqn. (4) and (5) should be vector $\vec{\phi}^\beta = (\langle m | \phi^\beta \rangle)_{m=1}^{2^n}$ because we are bounding each elements of $\vec{\phi}^\beta$ by its infinite norm, i.e: the largest element.)}

This extension of the localization landscape to many-body systems also places additional considerations on $\hat{H}$ for these bounds to hold, namely that sufficiently short-ranged hopping in $\hat{H}$ is required. 
For a Fock space graph $\mathcal{G}_F$ where nodes correspond to the $N$-spin states and edges connect state transitions according to the the hopping terms in the potential, this can be realized by maintaining the maximum degree of the Fock space graph $\mathcal{G}_F$ to be linear in $N$.

Further efforts in Ref.~\cite{filoche2021effective} explored the useful properties of the landscape function beyond disordered wave media, laying out additional constraints on the matrix form of $\hat{H}$ for these bounds to hold. More generally, $\hat{H}$ can be a positive semidefinite matrix with $\hat{H}_{ij} \leq 0$ for $i\neq j$, and $\hat{H}_{ij} \geq 0$ for $i=j$.

\section{Sampling from the landscape function}
\label{samplingll}

Our intention with this work is to prepare a quantum state $\ket{u}$ that represents the localization landscape function $u$, from which exact solutions to Eq.~\eqref{Eq:QUBO} can be sampled with probability $\alpha \left| \langle \vec{x}^* |u \rangle \right|^2$, where $\alpha$ is the number of degenerate solutions to the problem. 
Other low energy solutions can also be sampled with probabilities inversely proportional to their energy, as suggested by Eq.~\eqref{Eq:u_j}.

% In this section, we show how the landscape function $u$ can be used to obtain low energy solutions of $\hat{H}_{\textrm{Ising}}$ without having to prepare its ground state. 
% Our intention with this work is to prepare a quantum state $\ket{u}$ that represents the localization landscape function $u$, from which low energy solutions to Eq~\ref{Eq:QUBO} can be sampled with probabilities inversely proportional to their energy.
% Firstly, we show how a QUBO problem expressed in the form of $\hat{H}_{\textrm{Ising}}$ can be transformed to fit the constrains mentioned in \cite{filoche2021effective}.
% We outline a method to prepare a quantum state $\ket{u}$ representing the landscape function $u$.
% With the state $\ket{u}$, exact solutions to Eq~\ref{Eq:QUBO} can be sampled with probability $\alpha \left| \langle \vec{x}^* |u \rangle \right|$ where $\alpha$ is the number of degenerate solutions to the problem. 
% Other low energy solutions can also be sampled with probabilities inversely proportional to their energy, as suggested in Eq~\ref{Eq:u_j}.
% Sampling from $\ket{u}$ should allow for low energy states to be sampled with higher probability.
% Lastly, we investigate how the performance of our method scales with the two introduced hyperparameters $\Gamma$ and $\lambda$.

\subsection{Quadratic Unconstrained Binary Optimization}
\label{secQUBO}

The QUBO problem can be represented as
\begin{align}
    & \mathrm{Find } \; \; \vec{x}^* = \underset{\vec{x}}{\textrm{argmin}}\,\mathcal{C}_{\textrm{Q}}(\vec{x}) \label{Eq:argminQUBO} \\ 
    & \textrm{where } \mathcal{C}_{\textrm{Q}}(\vec{x}) =  \vec{x}^{\top} \mathcal{A} \vec{x}.
    \label{Eq:QUBO}
\end{align}
The vector $\vec{x}$ consists of $N$ binary variables, $\vec{x} = \left(x_1,...,x_N\right) \in \{0,1\}^N$, and $\mathcal{A}$ is a real symmetric matrix that defines the problem. 
Finding optimal solutions to QUBO problems, $\vec{x}^*$, is NP-hard in general~\cite{fu1986application}, and serves as a strong impetus for designing classical and quantum heuristics to find approximate solutions.
% Finding optimal solutions to QUBO problems, $\vec{x}^*$, is NP-hard in general~\cite{fu1986application}, so many classical and quantum heuristics have been designed to find approximate solutions. 

Quantum algorithms used to solve QUBO problems typically begin by mapping the QUBO cost function to an Ising Hamiltonian of the form,
\begin{equation}
\hat{H}_{\textrm{Ising}} =
\frac{1}{4} \sum_{ij}^N \mathcal{A}_{ij} (\hat{\sigma}^z_i + \hat{I}) (\hat{\sigma}^z_j + \hat{I})
\label{Eq:ising}
\end{equation} 
where $\hat{\sigma}^z_i$ is the Pauli-Z operator acting on the $i^{\textrm{th}}$ qubit. 
By mapping each binary variable in $\vec{x}$ to a qubit, the expectation value $\expval{\hat{H}_{\textrm{Ising}}}$ has a minimum value of $\mathcal{C}_{\textrm{Q}}(\vec{x}^*)$ in Eq.~\ref{Eq:QUBO}, and the QUBO problem can be solved by finding the ground state, $\ket{\vec{x}^*}$, that minimizes $\expval{\hat{H}_{\textrm{Ising}}}$.

\subsection{Fitting the constraints}
\label{constraints}

\begin{figure}
\begin{tabular}{cc}
\includegraphics[width=0.48\columnwidth]{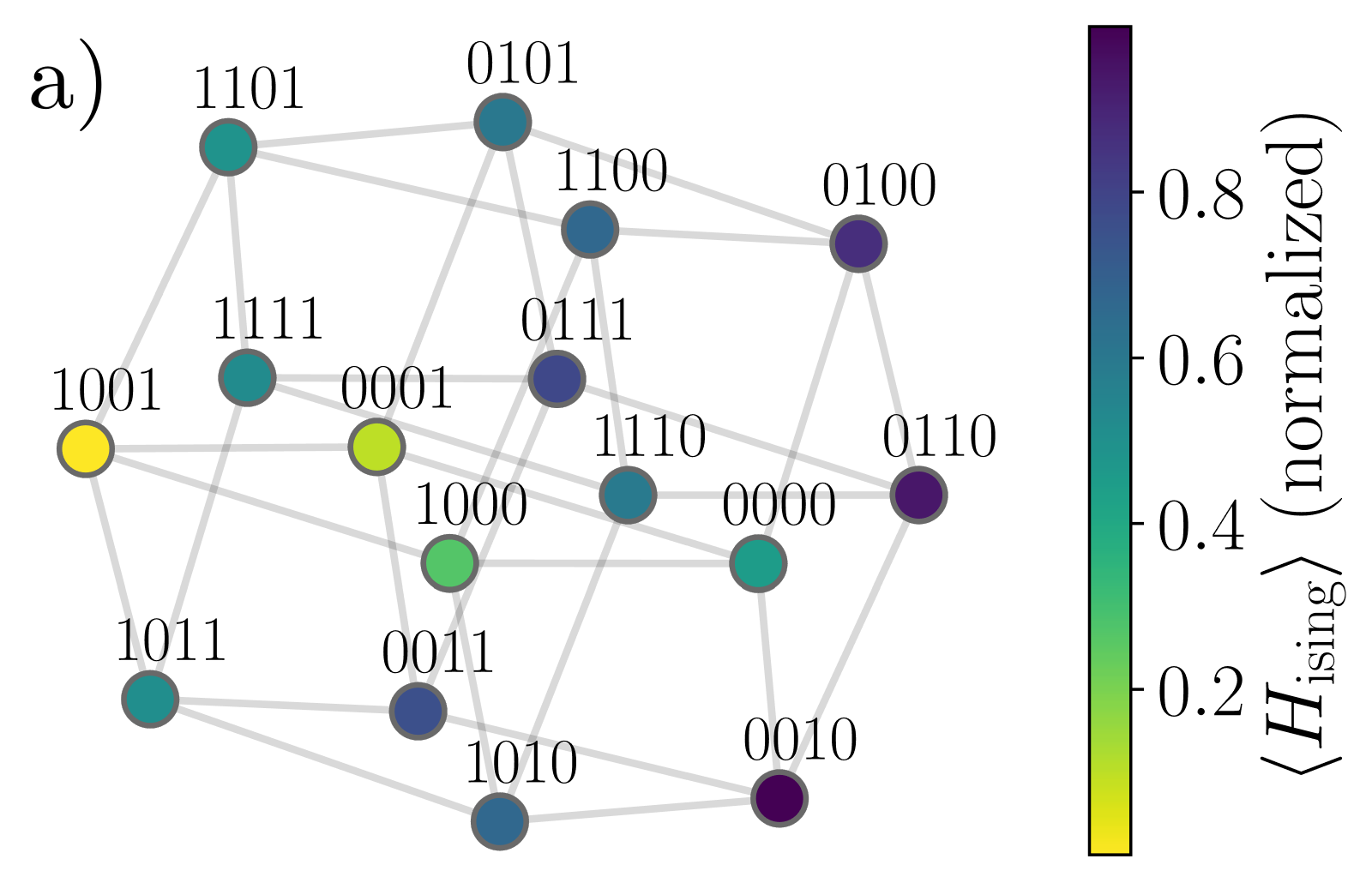} &
\includegraphics[width=0.48\columnwidth]{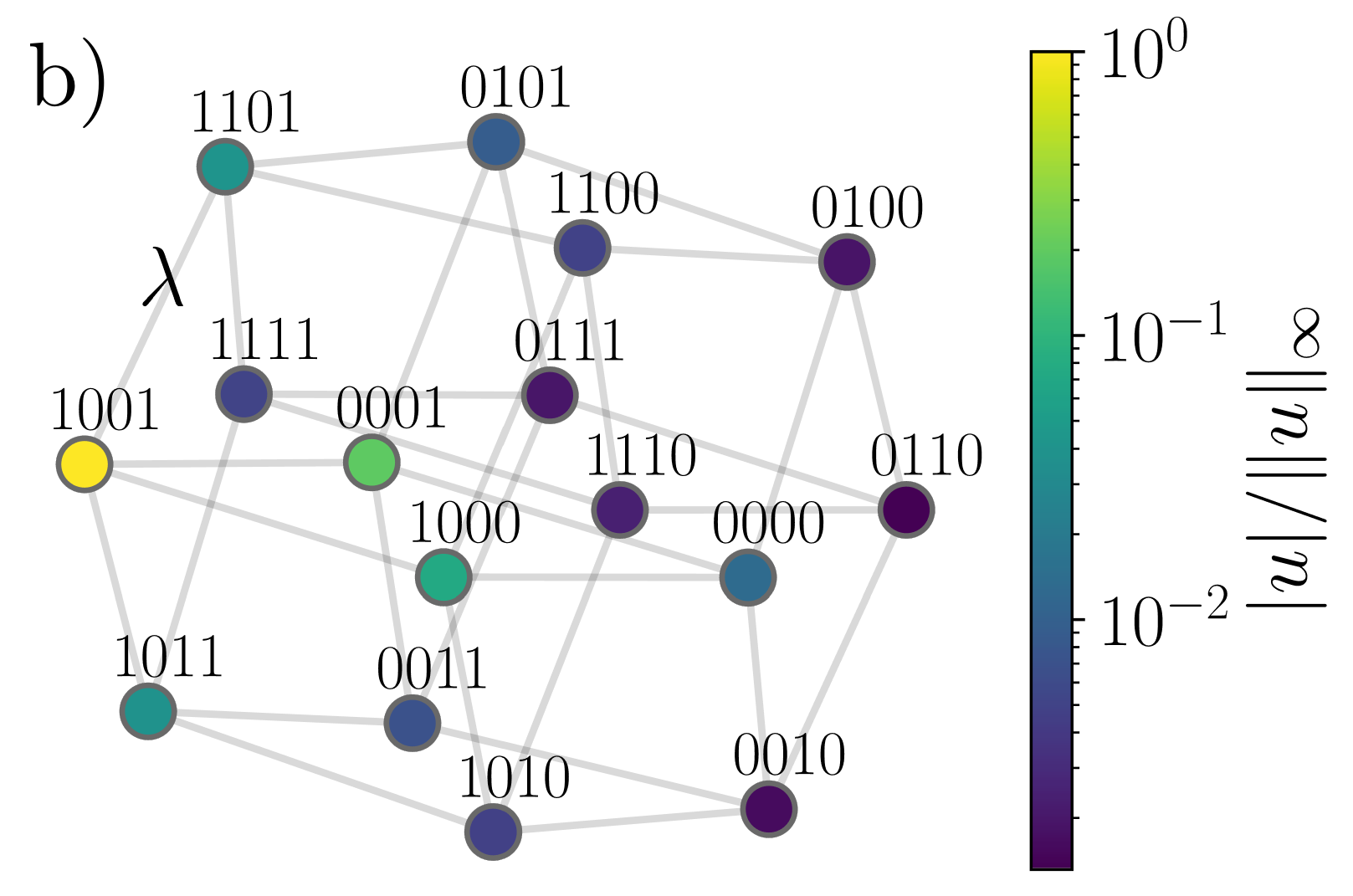}
\end{tabular}
    \caption{Representation of the modified Hamiltonian in Eq.~\eqref{Eq:hperturbed} as a graph in Fock space for an $N=4$ random QUBO instance. The considered perturbation induces an $N$-dimensional hypergraph structure with each bitstring coupled to $N$ nearest neighbours obtained by flipping one bit. Colours show a) values of 
    $\expval{\hat{H}_{\textrm{Ising}}}$ at each site normalized between $0$ and $1$, compared with b) the amplitude of the landscape function $\left|u\right| / \|u\|_{\infty}$ (right).}
    \label{fig:fock_graph}
\end{figure}

In general, $\hat{H}_{\textrm{Ising}}$ in Eq.~\eqref{Eq:ising} does not satisfy the aforementioned constraints for the landscape Eq.~\eqref{Eq:LLbounds} to bound the support of the low energy eigenstates.
However, the constraints can be satisfied by introducing the following transformation accompanied by two hyperparameters $\Gamma$ and $\lambda$:
\begin{equation}
\hat{H} = \hat{H}_{\textrm{Ising}} + \Gamma \hat{I} - \lambda \sum_i^N \hat{\sigma}_i^x
\label{Eq:hperturbed}
\end{equation} 
where $\hat{I}$ is the identity matrix.

The role of $\Gamma$ is to add a positive offset to the diagonal elements of $\hat{H}_{\textrm{Ising}}$ that is at least as large as its largest negative eigenvalue.
However, the largest negative eigenvalue is typically not known \textit{a priori} as it requires finding the solution to Eq.~\eqref{Eq:QUBO}, although in practice it is adequate to pick a sufficiently large value heuristically which can then be further fine tuned.

The ground state of $\hat{H}_{\textrm{Ising}}$ in Eq.~\eqref{Eq:ising} is a basis state in the computational $Z$-basis.
For problems with symmetries, such as the $\mathbb{Z}_2$ symmetry in MaxCut problems \cite{Bravyi_2020}, finding the exact ground state can lead to further ground states with the same energy.
In general, being able to find the ground state or an approximate ground state provides little information on nearby states with similar energy values, although there are heuristics that attempt to find ``nearby'' solutions in terms of energy \cite{goh2022techniques}.

The role of $\lambda$ is to introduce a mixing parameter into $\hat{H}_{\textrm{Ising}}$ to increase the overlap between states that are similar in terms of energy levels.
This is done so that the ground state of $\hat{H}$ in Eq.~\eqref{Eq:hperturbed} will contain components of surrounding low energy eigenstates of $\hat{H}_{\textrm{Ising}}$.
It is worth noting that by parameterizing $\lambda = \lambda(t)$, Eq.~\eqref{Eq:hperturbed} is often used as the Hamiltonian in Quantum Annealing, where one starts in the ground state of an easy-to-solve Hamiltonian in the large $\lambda$ limit and adiabatically decreases $\lambda(t)$ to zero to obtain the ground state of $H_{\mathrm{Ising}}$. 
The conditions imposed on $\hat{H}$ at the end of Sec.~\ref{LL} and the negative sign in Eq.~\eqref{Eq:hperturbed} limits $\lambda > 0$.

The Hamiltonian $\hat{H}$ can be visualized using a Fock space graph, $\mathcal{G}_F$, where nodes representing states of $\hat{H}_{\textrm{Ising}}$ are connected by an edge if they are one spin flip away, corresponding to the potential term $\lambda \sum_i^N \hat{\sigma}_i^x$ in Eq.~\eqref{Eq:hperturbed}.
An example of $\mathcal{G}_F$ for a $N=4$ Hamiltonian with randomly generated $\hat{H}_{\textrm{Ising}}$ with randomly chosen $\Gamma$ and $\lambda$ values satisfying these criteria is shown in Fig.~\ref{fig:fock_graph},
The similarities between the peak amplitudes of the localization landscape and the low energy states of $\hat{H}_{\textrm{Ising}}$ at each site can be observed, along with their decay based on the Hamming distance to the optimal solution (although the rate of decay is different).
The short-ranged hopping condition for the many-body localization landscape outlined at the end of Sec.~\ref{LL} is satisfied by $\mathcal{G}_F$ having a maximum degree of $N$.

Thus, we have shown that the QUBO problem can be mapped to a Hamiltonian whose low energy eigenstates are bounded by the localization landscape, at the cost of introducing two hyperparameters $\Gamma$ and $\lambda$, which control the tightness and extent in Hilbert space of the bounds provided by the landscape, respectively. While the process of finding the optimal values of $\lambda$ and $\Gamma$ for each problem instance is beyond the scope of this work, we will show some results on how they can affect the probability of sampling the optimal solutions in Sec.~\ref{hyperparameters}.

\begin{figure}
\begin{tabular}{c}
\includegraphics[width=0.95\columnwidth]{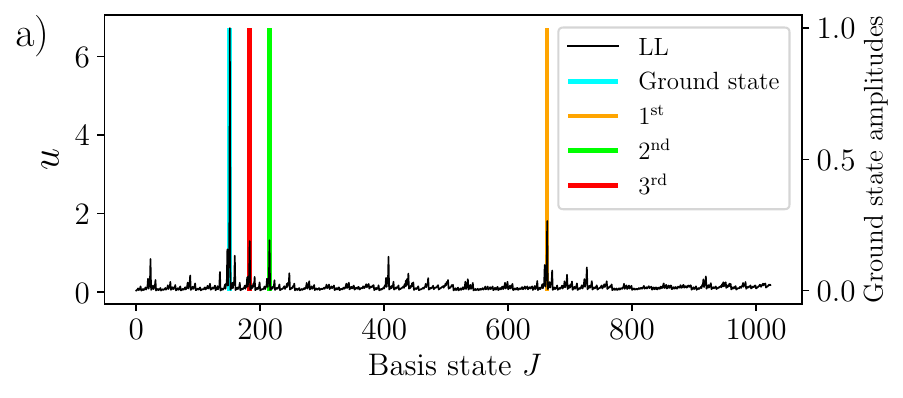} \\
\includegraphics[width=0.95\columnwidth]{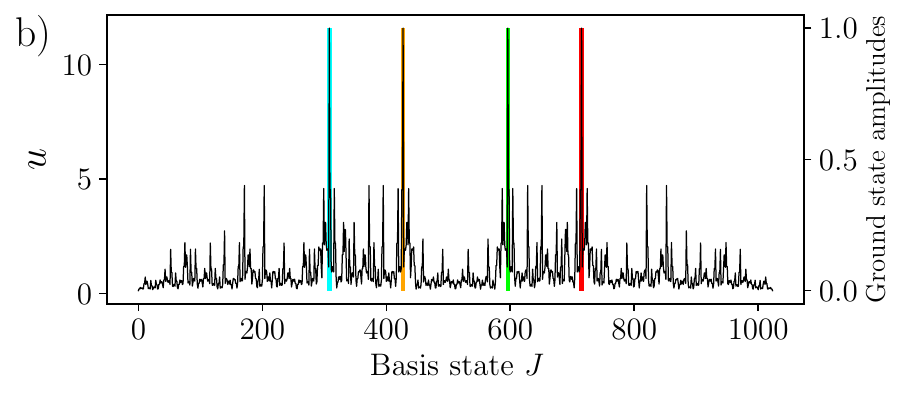}
\end{tabular}
    \caption{Localization landscape, $u$, of $\hat{H}$ for $N=10$ qubits as constructed in Eq.~\eqref{Eq:hperturbed}, compared with the $4$ lowest energy states of $\hat{H}_{\textrm{Ising}}$ for a) a randomly generated QUBO instance (non-degenerate case) and b) a randomly generated 3-regular MaxCut problem (degenerate case). The peaks of the landscape function correspond to the low energy eigenstates of $\hat{H}_{\textrm{Ising}}$. 
    }
    \label{fig:LLvsGS}
\end{figure}

\begin{figure*}
\centering  

\begin{tabular}{cc}
\includegraphics[width=0.45\linewidth]{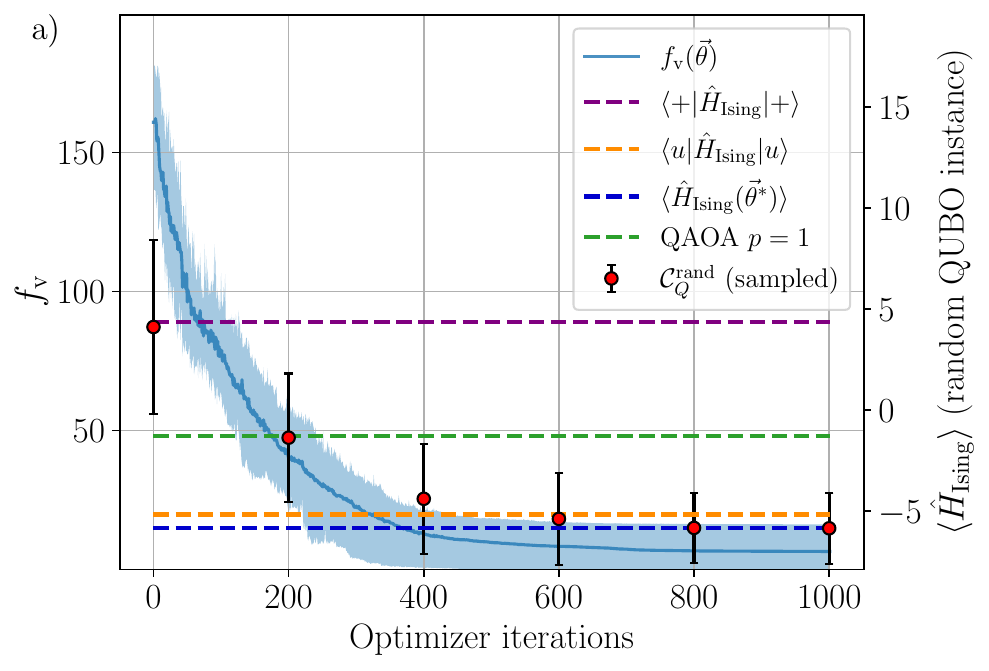} & 
\includegraphics[width=0.45\linewidth]{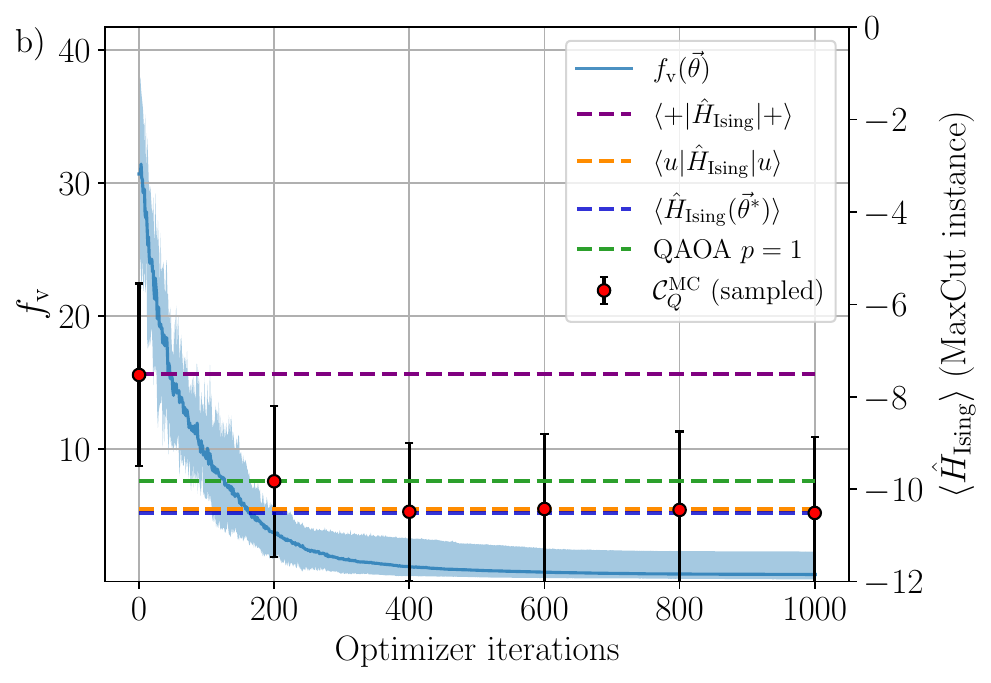}
\end{tabular}
    \caption{Variational search for $\ket{u}$ for a) a randomly generated QUBO problem with $-1<\mathcal{A}_{ij}<1$ (non-degenerate case) and b) a randomly generated 3-regular MaxCut problem (degenerate case).
    The variational search was performed using COBYLA to minimize $f_{\textrm{v}}(\vec{\theta})$ in Eq.~\eqref{eq:optC} over 10 unique initial sets of $\vec{\theta}$ using the statevector simulator in Pennylane. Solid lines show the average cost function at each optimizer iteration over 10 unique initial sets of $\vec{\theta}$. Shaded areas show the minimum and maximum of $f_{\textrm{v}}(\vec{\theta})$  over the 10 runs at each iteration.
    Dashed lines show different $\expval{\hat{H}_{\textrm{Ising}}}$ values, obtained from randomly sampling bitstrings over a uniform distribution (purple), preparing $\ket{u}$ exactly (orange), from the output state of our variational circuit after attempting to minimize $f_{\textrm{v}}(\vec{\theta})$ (blue), and from the optimized state of the QAOA with $p=1$ (green).
    Red markers and error bars show the average and standard deviation of classical QUBO cost function $\mathcal{C}_{\textrm{Q}}$ obtained from 10 bitstrings sampled every 200 iterations per optimization run.
    }
    \label{fig:ResultsLL}
\end{figure*}

\subsection{Preparing the landscape function}
\label{preparation}

Once we have the transformed Hamiltonian $\hat{H}$ the final step is to prepare $\ket{u}$, the state that corresponds to the landscape function of $\hat{H}$. Then, measuring $\ket{u}$ in the computational basis will sample bitstrings corresponding to peaks of $\ket{u}$ (or equivalently, valleys of the landscape) with higher probability. $\ket{u}$ can be obtained by solving the qubit-analogue of Eq.~\ref{Eq:LL} as a linear system of equations using a quantum device:
\begin{equation}
\hat{H} \ket{u} = \ket{+}
\label{Eq:qubitsLL}
\end{equation} 
where we use $\ket{+}$ to denote the $N$-qubit superposition state $\ket{+}^{\otimes N}$.

In this work, we use a variational method \cite{Cerezo_2021VQA} to prepare $\ket{u}$ using a variational ansatz $\ket{\psi(\vec{\theta})} = \hat{U}(\vec{\theta})\ket{0}$.
This is achieved by minimizing the following variational cost function:
\begin{equation}
    f_{\textrm{v}}(\vec{\theta}) = \left[\langle \psi (\vec{\theta}) | \hat{H} \ket{\psi (\vec{\theta})} - \langle \psi (\vec{\theta}) \ket{+}\right]^2
    \label{eq:optC}
\end{equation}
which is constructed from Eq.~\eqref{Eq:qubitsLL} by taking the inner product with $\ket{u}$ on both sides of the equation, squaring the difference between the two terms, and replacing $\ket{u}$ with a variational ansatz.
% \textcolor{cyan}{Probably more accurate to consider the squared-L2 norm just like in \cite{huang2021near} and rerun the results, although I doubt it will change much. Will require changing the gradient calculation (if relevant)}
% Depending on the choice of ansatz, minimizing $\mathcal{C}$ as a means of preparing $\ket{u}$ can be easily implemented on near term quantum devices.

We note that using a variational approach comes with several potential issues, namely the risk of encountering barren plateaus \cite{McClean_2018, Cerezo_2021, Wang_2021} or having limited expressibility where only a portion of the target state overlaps with states that can be produced. 
Notably, variational quantum algorithms can also be difficult to optimize \cite{Bittel_2021}.
% Despite the plethora of issues mentioned, we pursue the variational approach here for its simplicity as using a quantum device to solve linear systems of equations is not the main focus of our analysis.
Despite the plethora of issues, we pursue the variational approach here for its simplicity when implemented on NISQ devices and compatibility with shallow hardware-efficient circuits. 
Common techniques used to mitigate the effects of barren plateaus can also be applied \cite{liu2022mitigating, Liu_2023, Grant2019initialization, Mari2020transferlearningin, Skolik_2021}, although this was not required in obtaining the presented results. 
% \textbf{(BY: Can mention that all the BP mitigation methods can be applied to your scheme to increase the credibility of the method.)} \textcolor{cyan}{ah that's a good one.}

We contrast our variational approach here to traditional variational quantum approaches to QUBO problems, where the ground state of $\expval{\hat{H}_{\textrm{Ising}}}$ is typically not known, and the variational method is used to search for the optimal state, as opposed to preparing a known state.
One major advantage of our method is that the target state in our case is known and $\ket{u}$ can be obtained using any of the existing approaches for solving linear equations of the form $A\ket{x} = \ket{b}$, such as the well known HHL algorithm \cite{Harrow_2009}. 
Other, more NISQ-friendly methods for solving linear equations include variational methods such as the Variational Quantum Linear Solver (VQLS) \cite{bravoprieto2020variational, Patil_2022}, the Classical Combination of Variational Quantum States (CQS) \cite{huang2021near}, and the Hybrid Classical-Quantum Linear Solver \cite{Chen_2019}.
Regardless, the choice of method used to prepare $\ket{u}$ does not affect the validity of the following results.

% \textcolor{cyan}{[Need to discuss somewhere (here, intro, or conclusion) difference compared to standard variational algorithms, where the optimal state is typically unknown. Here on the other hand the target state is known (and could be obtained using other methods, say HHL), and the variational circuit is used to produce this state using a shallow circuit. So barren plateaus may be less of an issue here.]}

\section{Results}
\label{optresults}

% To demonstrate the effectiveness of using the landscape function to solve QUBO problems and our variational approach to prepare $\ket{u}$, we apply the methods described above to two problem instances with $N=10$ variables --- a randomly generated fully-connected QUBO matrix $\mathcal{A}$, with $\mathcal{A}_{ij} \in [-1,1]$ and a minimum of $\mathcal{C}^{\textrm{rand}}_{Q}(\vec{x}^*) \approx -7.895$ to showcase the non-degenerate case, and a randomly generated 3-regular MaxCut problem (formulated as a minimization problem) with a minimum of $\mathcal{C}^{\textrm{MC}}_{Q}(\vec{x}^*) = -12$ as a common example of a problem with multiple degenerate solutions.
To demonstrate the effectiveness of using the landscape function to solve QUBO problems and our variational approach to prepare $\ket{u}$, we apply the methods described above to two problem instances with $N=10$ variables --- a randomly generated fully-connected QUBO matrix $\mathcal{A}$, with $\mathcal{A}_{ij} \in [-1,1]$ to showcase the non-degenerate case, and a randomly generated 3-regular MaxCut problem (formulated as a minimization problem) as a common example of a problem with multiple degenerate solutions.
Using the landscape approximation for QUBO is generally problem agnostic, and in later sections, we use the non-degenerate case to further investigate the behaviour of the landscape function, and the degenerate case as an example of how prior knowledge about a structured problem can be used to improve the quality of the solutions.
% {\bf DL: It would be useful to also note what is the energy bandwidth of the Ising Hamiltonian. Needed to judge how big a precision nearest integer and nearest tenth is compared to the range of energy eigenvalues in the next paragraph.}

% For the small problem sizes, the exact solutions can be solved exactly, and we show the effectiveness of "good" values of $\Gamma$ and $\lambda$ by picking $\Gamma$ as the largest whole number such that $\Gamma + \mathcal{C}_{\textrm{Q}}(\vec{x}^*)$ is as close to $0$ as possible, and $\lambda$ was chosen to be the largest \textit{tenth} possible such that $\hat{H}^{-1} > 0$. \textcolor{cyan}{Clarify: $\Gamma$ is chosen first, followed by $\lambda$. The latter is kept small enough that the inverse remains non-negative?}

For small problem sizes, the exact solutions can be obtained exactly, and the minimum QUBO cost function for our MaxCut and randomly generated instances are $\mathcal{C}^{\textrm{MC}}_{Q}(\vec{x}^*) = -12$ and $\mathcal{C}^{\textrm{rand}}_{Q}(\vec{x}^*) \approx -7.895$ respectively.
To fit our problems to the constraints, we used a value of $\Gamma=8.5$ and $\lambda=0.3$ for the randomly generated QUBO instance, and $\Gamma=13$ and $\lambda=0.3$ for the MaxCut instance.
% To fit our problems to the constraints, we fix $\Gamma$ to be the smallest integer such that $\Gamma + \mathcal{C}_{\textrm{Q}}(\vec{x}^*)$ is positive, while being as close to $0$ as possible.
% We then chose $\lambda$ to be the largest value (up to a precision of 0.1) such that the inverse of $\hat{H}$ remains non-negative.

Figure~\ref{fig:LLvsGS} shows the landscape function $u$ of the respective perturbed Hamiltonians $\hat{H}$ for both the randomly generated QUBO instance and the MaxCut problem, compared with the $4$ lowest energy eigenstates for the Ising Hamiltonians representing the two problem instances.
We observe that the peaks of the landscape function $u$ line up with basis states of $\hat{H}$ which are the lowest energy eigenstates of $\hat{H}_{\textrm{Ising}}$.
Based off Fig.~\ref{fig:LLvsGS}, we intend to prepare the target state $\ket{u}$ that will have amplitudes of a similar structure to $u$ as shown in Fig~\ref{fig:LLvsGS}, from which these low energy states of $\hat{H}_{\textrm{Ising}}$ can be sampled with high probability.

% For each problem instance, we used the gradient-free classical optimizer COBYLA \cite{Powell1994, powell_1998, powell2007view} to search for the optimal parameters from $10$ initial starting set of $\vec{\theta}$ angles.
For both problem instances, we used the same variational ansatz consisting of an initial layer of Hadamard gates on all qubits, followed by $4$ alternating layers of $R_y(\theta)$ rotations on all qubits and nearest neighbour CNOT entangling gates in a linear topology, keeping all the coefficients of the quantum state real. 
We used COBYLA \cite{Powell1994, powell_1998, powell2007view}, a gradient-free classical optimizer, to search for the optimal parameters that minimizes Eq.~\ref{eq:optC} from $10$ initial starting set of $\vec{\theta}$ angles.
Gradient-based optimizers can also be used \cite{ruder2017overview}, and we show how the gradient of $f_{\textrm{v}}$ can be obtained in Appendix~\ref{appgradient} using a gate-based circuit that produces a quantum state with only real coefficients.

We compare the expectation value of $\expval{\hat{H}_{\textrm{Ising}}}$ obtained using our post-optimized state, $\ket{\psi(\vec{\theta}^*)}$, and from $\ket{u}$ obtained from inverting $\hat{H}$ in Eq.~\eqref{Eq:qubitsLL}.
We also compare the final $\expval{\hat{H}_{\textrm{Ising}}}$ values obtained using both our landscape method and from using the QAOA with $p=1$ layers. Finally, we compare the solutions obtained by sampling from our optimized ansatz, from $\ket{u}$, and from $\ket{\psi(\gamma, \beta)}_{\textrm{QAOA}}$ with $p=1$ as defined in Appendix~\ref{appqaoa}. 
All simulations were conducted using the statevector simulator (i.e. number of shots $\rightarrow \infty$) in PennyLane \cite{bergholm2022pennylane}.

In Fig.~\ref{fig:ResultsLL} we show the optimization runs used to prepare $\ket{u}$ using our variational ansatz, and the quality of the solutions obtained after every $200$ iterations of COBYLA used to calculate the classical QUBO cost function, $\mathcal{C}_{\textrm{Q}}$ in Eq.~\eqref{Eq:QUBO} for both the MaxCut and random QUBO instances.
Also shown in Fig.~\ref{fig:ResultsLL} are comparisons between $\expval{\hat{H}_{\textrm{Ising}}}$ from preparing the exact $\ket{u}$, from randomly sampling bitstrings over a uniform distribution, from optimizing the QAOA with $p=1$, and from our variational ansatz after optimization $\langle \hat{H}_{\textrm{Ising}} (\vec{\theta}^*) \rangle$ = $\langle \psi(\vec{\theta}^*) | \hat{H}_{\textrm{Ising}} | \psi(\vec{\theta}^*) \rangle$.
Further information regarding our implementation of the QAOA and the number of CNOT gates used can be found in Appendix~\ref{appqaoa}.

As observed in Fig.~\ref{fig:ResultsLL}, the mean QUBO cost function from sampled bitstrings obtained every $200$ iterations tend towards $\langle u |\hat{H}_{\textrm{Ising}}| u \rangle$ as the ansatz converges to a state representing $\ket{u}$.
%In both problem instances, being able to prepare the exact $\ket{u}$ allows one to obtain a lower cost function value compared with the QAOA with our simple circuit ansatz, while using much fewer gates than the number required to decompose the QAOA ansatz.
In both problem instances, being able to prepare $\ket{u}$, whether exactly or using our simple circuit ansatz, allows one to obtain a lower cost function value compared with the QAOA.

For the non-degenerate case, being able to prepare and sample from the exact landscape function state $\ket{u}$ brings us closer to the optimal $\mathcal{C}_{\textrm{Q}}$ value compared to $p=1$ of the QAOA. 
This is likely due to our specific problem and choice of hyperparameters, where the mixing introduced by $\lambda$ in Eq.~\ref{Eq:hperturbed} is small compared to the difference between the lowest two energy states of $\expval{\hat{H}_{\textrm{Ising}}}$ for the non-degenerate case.
This causes the ground state of $\expval{\hat{H}_{\textrm{Ising}}}$ to be the dominant basis state in the ground state of $\hat{H}$ and preparing $\ket{u}$ will produce a strong peak at $\ket{\vec{x}^*}$.

% {\bf DL: Note physical review style is to label figure sub-panels by (a), (b), etc. Also I suggest to use lower case $p$ for the QAOA order in line with the common notation used in the literature on QAOA - please amend figures accordingly}

\section{Effect of Hyperparameters}
\label{hyperparameters}

\begin{figure}
\begin{tabular}{cc}
\includegraphics[width=0.49\columnwidth]{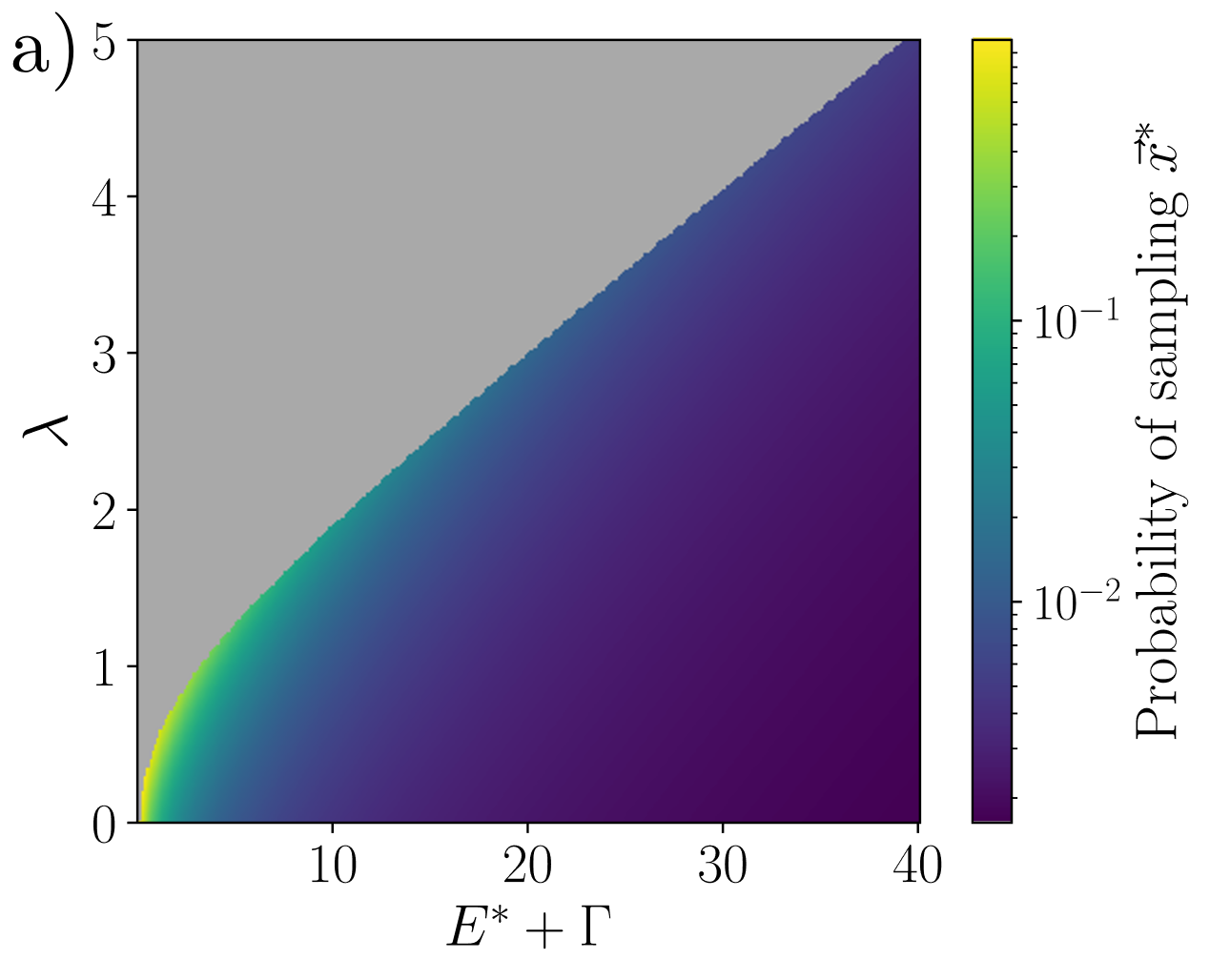} &
\includegraphics[width=0.49\columnwidth]{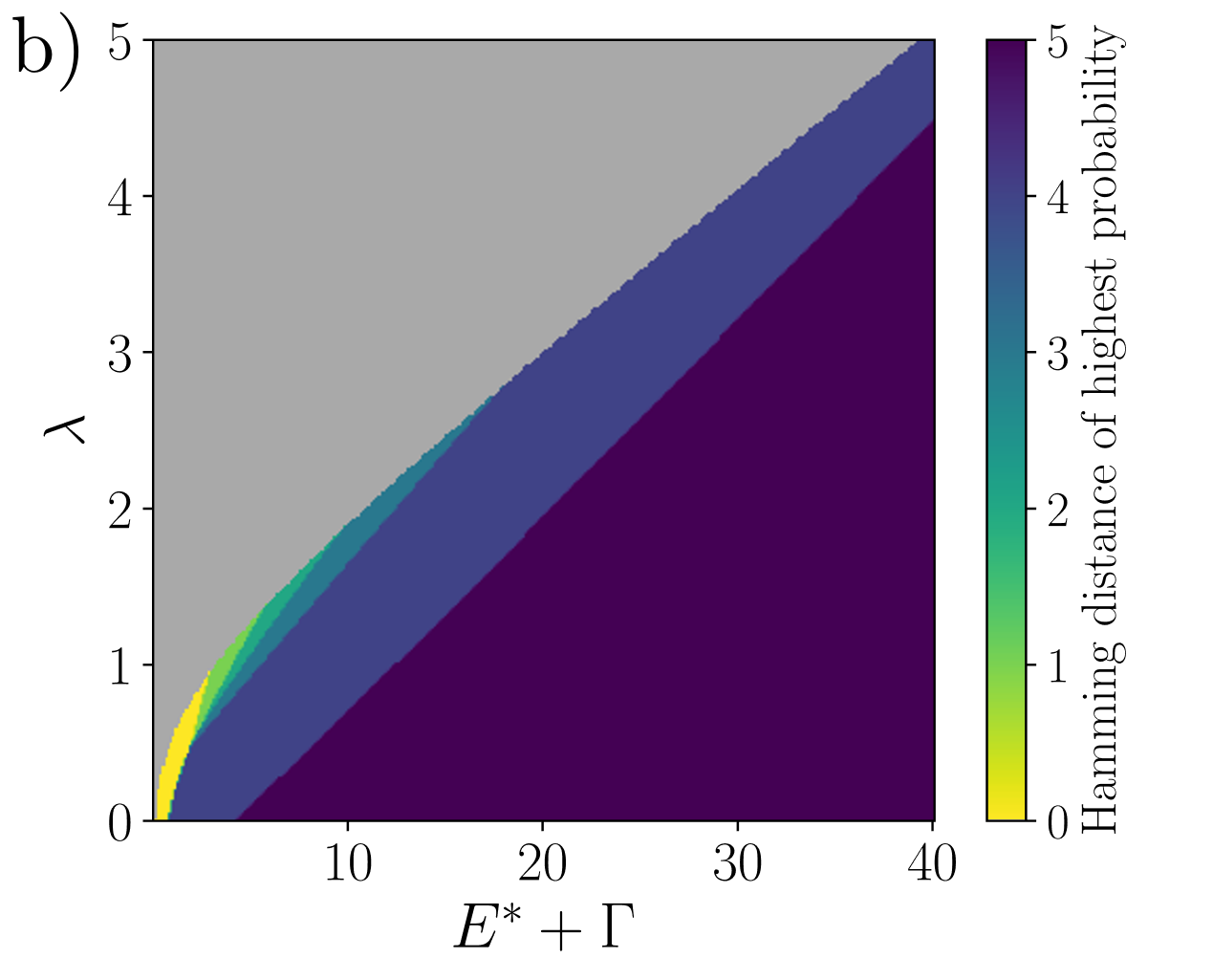}
\end{tabular}
    \caption{(a) Probability of sampling $x^*$ from $\ket{u}$ as a function of hyperparameters $\lambda$ and $\Gamma$ for the random QUBO instance. (b) Bitstrings with Hamming distance to solution most likely to be sampled from $\ket{u}$ as a function of hyperparameters $\lambda$ and $\Gamma$ for the random QUBO instance. Grey areas show values of $\lambda$ and $\Gamma$ for which $\hat{H}^{-1} < 0$, where the conditions for the landscape to bound the low energy eigenstates are not satisfied.}
    \label{fig:hyperparam}
\end{figure}

% In this section we explore how the hyperparameters $\Gamma$ and $\lambda$ affect the quality of the solutions obtained using some analysis on the non-degenerate case, although our numerical results show similar properties for degenerate problem instances as well.

In this section we explore how the hyperparameters $\Gamma$ and $\lambda$ affect the quality of the solutions obtained for the non-degenerate case, although similar properties hold for degenerate problem instances as well.

To properly characterize the capabilities of $\ket{u}$, the results presented from this section on are limited to states that can be prepared exactly.
As mentioned in Sec.~\ref{preparation}, the variational method in Sec.~\ref{optresults} was mainly an example of how $\ket{u}$ can be prepared quickly using NISQ-friendly methods, and other methods can be used to $\ket{u}$ with potentially higher accuracy. 

We begin by noting that the optimal solution to the QUBO problem in Eq.~\eqref{Eq:argminQUBO} can be represented by a computational basis state of $\hat{H}_{\textrm{Ising}}$ in Eq.~\eqref{Eq:ising} used to construct $\hat{H}$.
Using Eq.~\eqref{Eq:LLbounds}, we can find the probability amplitude associated with sampling $\ket{\vec{x}^*}$ if we had prepared the ground state of $\hat{H}$:
\begin{equation}
    u_{x^*} \left| E^\beta \right| \left\| \phi_\beta \right\|_\infty  \geq \left| \langle \vec{x}^* | \phi^\beta \rangle \right| 
    \label{Eq:samplingineq}\\
\end{equation} 
where in this case we let $\ket{\phi^\beta}$ and $E^\beta$ to be the ground state and ground state energy of $\hat{H}$, respectively.
For small values of $\lambda$, we can expand the denominator using first order perturbation theory and express $E^\beta$ in terms of $\Gamma$, $\lambda$, and the ground state energy of $\hat{H}_{\textrm{Ising}}$ (i.e. $E^* = \mathcal{C}_Q(\vec{x}^*)$).
\begin{align}
    u_{x^*} & \geq \frac{\left| \langle \vec{x}^* | \phi^\beta \rangle \right|}{\left| E^\beta \right| \left\| \phi_\beta \right\|_\infty} \label{Eq:ujbounds} \\
     & \approx \frac{\left| \langle \vec{x}^* | \phi^\beta \rangle \right|}{\left| E^* + \Gamma + \lambda \langle \vec{x}^*|\sum_i \hat{\sigma}_i^x| \vec{x}^*\rangle \right| \left\| \phi_\beta \right\|_\infty}  \label{Eq:ujperturb}
\end{align} 

\begin{figure}
\includegraphics[width=0.97\columnwidth]{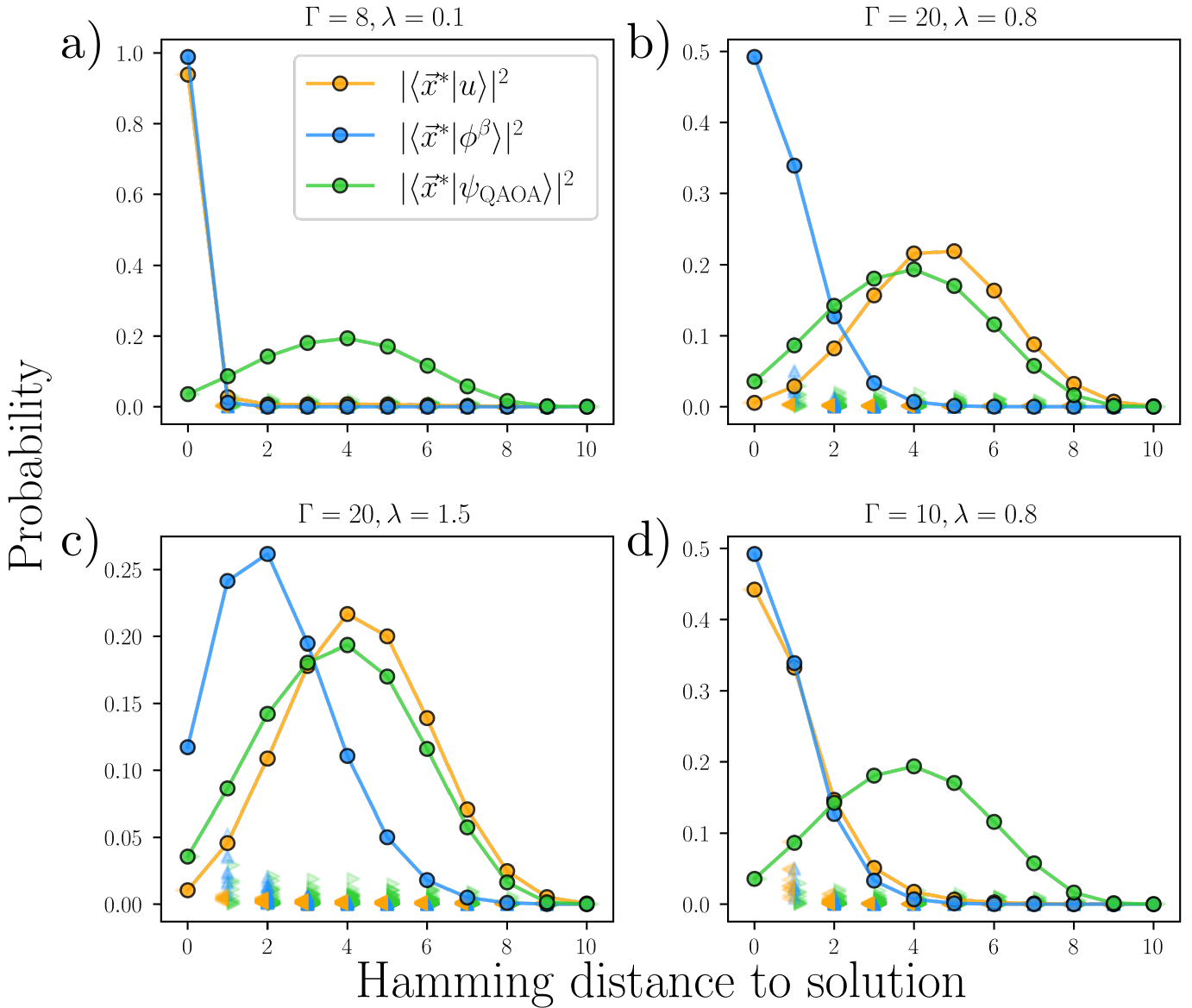} 
    \caption{Probability of sampling a solution with Hamming distance $d$ away from the optimal solution $\vec{x}^*$ for the randomly generated QUBO problem for different values of $\Gamma$ and $\lambda$. Coloured triangles show the probability of sampling these individual solutions from $\ket{u}$ (orange), the ground state of $\hat{H}$ (blue), and from the optimal state of the QAOA with $p=1$ (green). Connected lines show the total probability of all solutions of Hamming distance $d$ away from the optimal solution (i.e. the sum of all the triangles at a given $d$). 
    %\textbf{(BY: Would it be clearer if we set the limits for y-axis for all graph to be the same? But keep this if the graph looks ugly and not informative after changing the limits.}
    % I personally would keep it as is ba, cause part of graph reading is to look at the axes and scale lol. 
    } 
    \label{fig:hyperparamhd}
\end{figure}

The left hand side of Eq.~\eqref{Eq:ujbounds}, $u_{x^*}$, is not $| \langle \vec{x}^* | u \rangle | $, since the landscape function $u$ from Eq.~\eqref{Eq:LLbounds} is not a normalized state. Nevertheless, it is related to the probability amplitude of sampling $\ket{\vec{x}^*}$ from $\ket{u}$ and it is still in our interest to maximize it.
According to Eq.~\eqref{Eq:ujperturb}, this can be done by choosing $\Gamma$ to be as close to $-E^*$ as possible.

Figure~\ref{fig:hyperparam}(a) shows the probability of sampling the optimal solution, $\vec{x}^*$ from $\ket{u}$ as a function of both $\Gamma$ and $\lambda$. 
Values of $\lambda > 1$ are beyond the perturbative regime used in the approximation in Eq.~\eqref{Eq:ujperturb}.

Shown in Fig.~\ref{fig:hyperparam}(b) is the Hamming distance between the optimal solution $\vec{x}^*$ and the vector $\vec{x}$ that has the highest probability of being sampled from $\ket{u}$. This can be expressed more succinctly as
\begin{equation}
    d(\vec{x}^*, \underset{\vec{x}}{\textrm{argmax}} \left|\langle \vec{x} | u \rangle \right|^2 ),
\end{equation}
where $d(\vec{x}_1,\vec{x}_2)$ is the Hamming distance between $\vec{x}_1$ and $\vec{x}_2$. 
The results in Fig.~\ref{fig:hyperparam} also suggest that sampling solutions close to the optimum favour having $\lambda$ to be as large as possible while still respecting the constraints for a given $E^* + \Gamma$, which should be as close to $0$ as possible. 

In Fig.~\ref{fig:hyperparamhd}, we compare the total probability of sampling solutions with Hamming distance $d$ away from the optimal solution for the randomly generated QUBO problem from $3$ states --- $\ket{u}$ , the ground state of the perturbed Hamiltonian $\hat{H}$, and the same optimal state of the QAOA with $p=1$ in Fig.~\ref{fig:ResultsLL}.
These probabilities change as a function of $\Gamma$ and $\lambda$.

In Fig.~\ref{fig:hyperparamhd}(a) $\Gamma$ and $\lambda$ are too small, and there is insufficient mixing between the low energy states of the corresponding $\hat{H}_{\textrm{Ising}}$, and preparing the landscape function provides similar probabilities to sample the grounds state of $\hat{H}_{\textrm{Ising}}$ from the perturbed Hamiltonian. 
This can be desirable in most cases where we are only interested in the optimal solution to the QUBO problem.

In contrast, for cases where $\Gamma$ and/or $\lambda$ are too large, such as in Fig.~\ref{fig:hyperparamhd}(b,c), the majority of the solutions sampled will be approximately $\frac{N}{2}$ Hamming distances away. 
For large values of $\lambda$, the perturbation term in $\hat{H}$ become dominant compared to the $ZZ$-interactions in $\hat{H}_{\textrm{Ising}}$, and $\ket{u}$ tends toward the uniform superposition state in the computational basis. 

% . \textbf{(BY: I think we should also highlight the result of QAOA is not that good as the probability of sampling $\frac{N}{2}$ hamming distance away is dominated for QAOA.)}
% % Ben: Added

However, there is also an interesting regime in Fig.~\ref{fig:hyperparamhd}(d) where, for well-chosen values of $\Gamma$ and $\lambda$, sampling from $\ket{u}$ is able to provide the optimal solution with a high probability along with nearby solutions in terms of Hamming distance. 
In practice, this can be used to find the optimal bitstring from just a handful of samples on $\ket{u}$.

On the other hand, sampling from the optimal state produced by the QAOA with $p=1$ will result in majority of the samples being $\frac{N}{2}$ hamming distance away from the optimal solution.

We note that in all of these cases, the probability of obtaining the optimal solution $\vec{x}^*$ from $\ket{u}$ is still higher than any other bitstring, although it may not form the majority of the samples obtained.

% The LHS of Eq.~\ref{Eq:ujbounds}, $u_{x^*}$, while related to the probability amplitude of sampling $\ket{\vec{x}^*}$ from $\ket{u}$, is not $| \langle \vec{x}^* | u \rangle | $ as the landscape function $u$ from Eq.~\ref{Eq:LLbounds} is not a normalized state. 

% The LHS of Eq.~\ref{Eq:ujbounds}, $u_{x^*}$, while related to the probability amplitude of sampling $\ket{\vec{x}^*}$ from $\ket{u}$, is not $| \langle \vec{x}^* | u \rangle | $ as the landscape function $u$ from Eq.~\ref{Eq:LLbounds} is not a normalized state. 
% However, Eq.~\ref{Eq:ujperturb} is still able to provide some insights into choosing $\Gamma$. 
% The quantity $E^*$ in the RHS is fixed by the QUBO problem and the numerator $\left| \langle \vec{x}^* | \phi^\beta \rangle \right|$ does not vary with $\Gamma$ as it adds a fixed offset to the Hamiltonian $\hat{H}$.
% Additionally, it is easy to check that the term linear in $\lambda$ in the denominator evaluates to 0. 
% To increase the component in $u$ that is relevant to the peak of $\ket{\vec{x}^*}$, we would require $|E^* + \Gamma|$ to be as small as possible.
% For QUBO problems where the optimal cost function is positive, this also means that we want $\Gamma$ to be a negative offset such that the smallest element in $\hat{H}$ is positive and close to $0$.

\begin{figure}
\begin{tabular}{c}
\includegraphics[width=0.95\columnwidth]{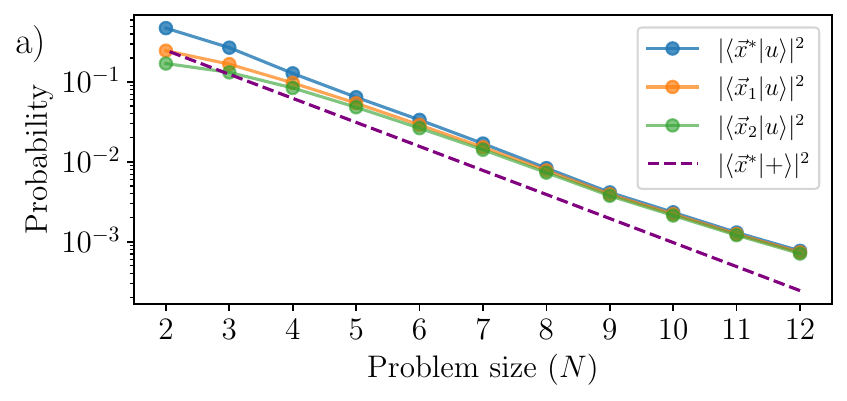} \\
\includegraphics[width=0.95\columnwidth]{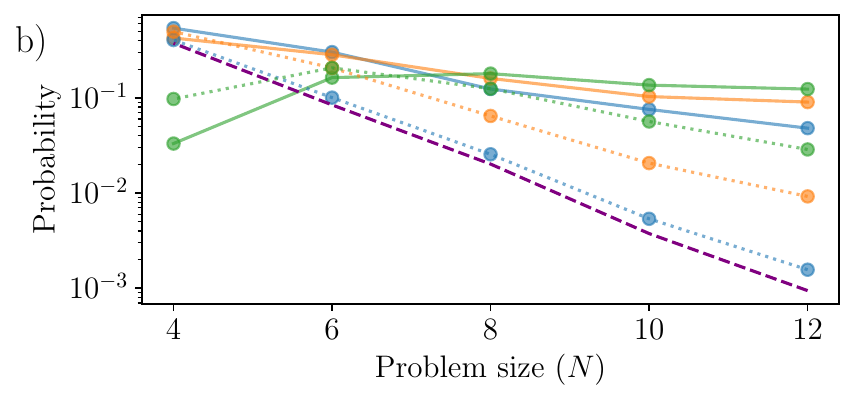}
\end{tabular}
    \caption{Probability of sampling eigenstates of $\hat{H}_{\textrm{Ising}}$ corresponding to the first $3$ lowest energy states from $\ket{u}$ with for (a) randomly generated QUBO problems (non-degenerate case) and (b) randomly generated $3$-regular MaxCut problems (degenerate case) . 
    Solid lines in (b) show these probabilities using $\Gamma$ and $\lambda$ values chosen with some prior knowledge of the problem instances.
    Dotted lines in (b) show these probabilities using the same method of choosing $\Gamma$ and $\lambda$ values as in (a). 
    Purple dashed lines shows the probability of sampling $\vec{x}^*$ from a uniform distribution.
    % For the MaxCut problem, this is the total probability of sampling any of the degenerate solutions with the same energy as the ground state, and the first and second excited states.
    Each plot point was obtained by averaging over $100$ randomly generated problem instances, and $\ket{u}$ was found by solving Eq.~\eqref{Eq:qubitsLL} exactly.
    }
    \label{fig:LLvsN}
\end{figure}

The main results presented so far mainly concerned $2$ different problem instances for $N=10$.
Figure~\ref{fig:LLvsN} represents an initial foray into how using the localization landscape scales with problem sizes, as well as how prior knowledge of the problem can be used to increase the probability of sampling optimal solutions.

For each problem type (random QUBO and MaxCut instances), the probability of sampling eigenstates of $\hat{H}_{\textrm{Ising}}$ ($\vec{x}^*$, $\vec{x}_1$, and $\vec{x}_2$) corresponding to the $3$ distinct lowest energy values ($E^*$, $E_1$, and $E_2$) are plotted against the problem size $N$, and compared against the probability obtained from sampling the optimal solution from a uniform distribution. 
For the MaxCut instances in Fig.~\ref{fig:LLvsN}(b), the curves show the total probability, i.e. $\alpha \left| \langle \vec{x}_i | u \rangle \right|^2$, where $\alpha$ is the number of degenerate states corresponding to energy $E_i$.

At first glance, it is worth noting that while the exponentially decreasing probability of sampling the optimal solution may pose a glaring issue, especially for the random QUBO instance, the probability of sampling $\vec{x}^*$ remains consistently above that of random sampling.

As mentioned earlier, a ``good" choice of $\Gamma$ would be one that is as close to $-E^*$ as possible. 
For any QUBO problem, $-\sum_{i,j}\left|\mathcal{A}_{ij}\right|$ is a lower bound of $\mathcal{C}_Q$ and an initial value $\Gamma = 1.1\times \sum_{i,j}\left|\mathcal{A}_{ij}\right|$ can be used for unstructured problems.

For a MaxCut problem, one can use the maximum possible number of edge bisections in a graph as the lower bound for $\mathcal{C}_Q$. 
This is equal to the total number of edges, $n_e = \frac{Nd}{2}$, for a $d$-regular graph. 
For $3$-regular graphs, one can choose $\Gamma = \frac{3N}{2} + 1 $ which is typically less than $ \sum_{i,j}\left|\mathcal{A}_{ij}\right|$ to obtain a much higher probability in sampling $\vec{x}^*$, as observed when comparing the solid and dotted lines in Fig.~\ref{fig:LLvsN}(b).

We used a value of $\lambda=0.07\:\Gamma$ for the random QUBO instances, and $\lambda=0.03\:\Gamma$ for the MaxCut problems to fulfill the constraints in Sec.~\ref{LL} for the instances considered in Fig.~\ref{fig:LLvsN}.
However, these values may not be valid or ideal for all QUBO problems in general. 
In all problem instances, the exponential decrease in probability can be ameliorated with further tuning of $\Gamma$ and $\lambda$ for the specific instance.

Another interesting observation of Fig.~\ref{fig:LLvsN}(b) is how higher energy states can have a greater overall probability of being sampled compared to the optimal solution. 
This can be explained by the increase in number of degenerate states closer to the middle of the energy spectrum. 
As shown in Fig.~\ref{fig:LLvsGS}(b), the probability of sampling individual ground states is still dominant compared to the other states.

\section{Discussion and conclusion}
\label{conclusion}

A key part in obtaining the results in this work was by perturbing $\expval{\hat{H}_{\textrm{Ising}}}$, which is diagonal in the computational basis, with a uniform transverse magnetic field $\sum_i^N \hat{\sigma}_i^z$, equivalent to a uniform nearest neighbour hopping on an $N$-dimensional hypercube.
This was done to controllably smear out the eigenstates of $\hat{H}_{\textrm{Ising}}$ in the Fock space, allowing for the QUBO problem to be solved approximately by sampling from the solutions of the easier-to-solve landscape problem.

As we have observed, the quality of the resulting solutions will depend on the strength and form of the perturbing potential, and the properties of alternative perturbation terms provide interesting avenues for further exploration. 
For example, one may replace the perturbative term with a number-conserving perturbation (arising in the case of models of many-body localization), such as $\sum_{i,j} \left(\hat{\sigma}_i^+ \hat{\sigma}_j^- + \textrm{h.c.} \right)$ where $\hat{\sigma}^{\pm} = \frac{1}{2} (\hat{\sigma}^{x} \pm i \hat{\sigma}^{y})$, leading to landscape functions that explore decoupled subspaces of the full Hilbert space as in Ref.~\cite{PhysRevB.101.014201}.
Such a perturbative term may be more useful when sampling solutions to QUBO problems involving hard constraints, such as those requiring the number of spin excitations to be preserved.

%\textcolor{cyan}{I'm actually quite curious as to how this would work (if possible). How can one select beforehand whether they want to sample states with a specific number of spin-ups so that they can prepare the LL to efficiently sample from these states?}

%{\bf DL: This would work by modifying the perturbation term proportional to $\lambda$. In our current treatment, the perturbation allows for any spin to be flipped, meaning that the eigenstates of $\hat{H}_{Ising}$ get smeared out over the whole Hilbert space (of course, spreading between bitstrings with similar energies is favoured, as follows from perturbation theory arguments). But one could just as easily consider a perturbation of the form $\lambda \sum_{\langle ij \rangle} \hat{\sigma}_i^x \hat{\sigma}_j^x$, which keep the total number of up spins conserved. With such a perturbation, Hamiltonian is no longer a simple diagonal matrix in the computational basis, but it block diagonal with sectors with different numbers of up spins remaining decoupled.}

Another interesting avenue for exploration would be to consider QUBO problems where the eigenvalue spectrum of the Hamiltonian encoding the problem is skewed towards having a few low energy states separated from many high energy states by a large gap. 
These types of QUBO problems are typically present in industry-relevant contexts, where the use of a penalty term when constructing the unconstrained problem causes all solutions that do not satisfy any constraints to have very high costs. 
By preparing the landscape function, it should be possible to prepare a state such that solutions satisfying the constraints can be sampled much easily, and the optimal solution can be easily found from this smaller, finite group of samples.

In conclusion, we showed how to apply the localization landscape theory used to find localized regions of low energy eigenstates in many-body systems to prepare quantum states that can be used to sample low energy solutions to the QUBO problem with high probability. We demonstrated our methods on two problem instances, a randomly generated MaxCut problem \cite{garey1974some, glover2019tutorial} exemplifying the degenerate case and a randomly generated QUBO problem for the non-degenerate case, and showed that by preparing a state, $\ket{u}$, representing the landscape function, low energy solutions to the Ising Hamiltonian corresponding to the QUBO problem can be sampled with higher probability. An advantage of the approach is that the good solutions can be sampled using relatively shallow circuits, minimizing the effect of gate noise and decoherence present in current noisy intermediate-scale quantum processors.

\section*{Acknowledgements} 

We acknowledge support from the National Research Foundation, Prime Minister’s Office, Singapore and A*STAR under the CQT Bridging Grant and the Quantum Engineering Programme (NRF2021-QEP2-02-P02), and by the EU HORIZON-Project101080085—QCFD.

% Further work to investigate:
% \begin{itemize}
%     \item Different forms for the potentials e.g. $\sum_{i,j} \hat{\sigma}_i^x \hat{\sigma}_j^x$ - analogous to choice of mixer Hamiltonian in QAOA or annealing path in QA
%     \item Also can consider potentials such as $\sum_{i,j} \hat{\sigma}_i^+ \hat{\sigma}_j^-$
%     \item Different kinds of QUBO problems with different energy level distribution
% \end{itemize}

% Some discussion on the choice of different forms for the potentials: a key part of the approach here was to perturb the Hamiltonian encoding the QUBO problem (diagonal in the computational basis / Fock space) to smear out its eigenstates in Fock space, allowing the QUBO problem to be solved (approximately) by sampling from solutions of an easier-to-solve problem (the landscape problem). The resulting solutions will depend on the strength and form of the perturbing potential; we considered here the case of a uniform transverse magnetic field, equivalent to a uniform nearest neighbour hopping on an $N$-dimensional hypercube. It might be interesting to explore the properties of other perturbation terms. For example, the introduction of a number-conserving perturbation (arising in the case of models of many-body localization) leads to landscape functions that explore decoupled subspaces of the full Hilbert space, see e.g. Ref.~\cite{PhysRevB.101.014201}. This may be useful for sampling from solutions of QUBO problems involving hard constraints.

\bibliography{main} % Produces the bibliography via BibTeX.

\appendix

\section{Gradient calculation}
\label{appgradient}

In this Appendix, we show how the derivative of the cost function, $\frac{\partial f_{\textrm{v}}}{\partial \theta_i}$, can be obtained in-situ using a quantum device and the parameter shift rule.
\begin{align}
    f_{\textrm{v}} = & \left[\langle \psi (\vec{\theta}) | \hat{H} \ket{\psi (\vec{\theta})} - \langle \psi (\vec{\theta}) \ket{+}\right]^2 \\
    \frac{\partial f_{\textrm{v}}}{\partial \theta_i} = & \; 2 \left[\langle \psi (\vec{\theta}) | \hat{H} \ket{\psi (\vec{\theta})} - \langle \psi (\vec{\theta}) \ket{+}\right]\\
    & \times \frac{\partial}{\partial \theta_i} \left[\langle \psi (\vec{\theta}) | \hat{H} \ket{\psi (\vec{\theta})} - \langle \psi (\vec{\theta}) \ket{+}\right]
\end{align}
From here, we will proceed term by term. Using the parameter shift rule:
% \begin{align}
%     \frac{\partial}{\partial \theta} \langle \psi (\vec{\theta}) | \hat{H} \ket{\psi (\vec{\theta})} = &
%     \frac{1}{2} \left[ 
%     \expval{\hat{H}}(\theta + \frac{\pi}{2}) \right. \nonumber\\
%     & - 
%     \left. \expval{\hat{H}}(\theta - \frac{\pi}{2}) 
%     \right]
% \end{align}
\begin{equation}
    \frac{\partial}{\partial \theta_i} \langle \psi (\vec{\theta}) | \hat{H} \ket{\psi (\vec{\theta})} = 
    \frac{1}{2} \left[ 
    \expval{\hat{H}}(\theta_i + \frac{\pi}{2})
     - 
    \expval{\hat{H}}(\theta_i - \frac{\pi}{2}) 
    \right]
\end{equation}
To evaluate $\frac{\partial}{\partial \theta_i} \left( \langle \psi (\vec{\theta}) \ket{+} \right)$, we observe that for a real quantum state $\ket{\psi ({\vec{\theta}})}$:
\begin{equation}
    \frac{\partial}{\partial \theta_i} | \langle \psi (\vec{\theta}) \ket{+} | ^2 
    =
    2 \langle \psi (\vec{\theta}) \ket{+} \frac{\partial}{\partial \theta_i} \left( \langle \psi (\vec{\theta}) \ket{+} \right)
    \label{eq:iden1}
\end{equation}
and 
\begin{equation}
    \frac{\partial}{\partial \theta_i} | \langle \psi (\vec{\theta}) \ket{+} | ^2 
    =
    \frac{\partial}{\partial \theta_i} \left( \langle \psi (\vec{\theta}) | + \rangle \langle + | \psi (\vec{\theta}) \right).
    \label{eq:iden2}
\end{equation}

Putting Eq.~\eqref{eq:iden1} and Eq.~\eqref{eq:iden2} together, and letting $\hat{M} = | + \rangle \langle + | = \left(\frac{I + \hat{\sigma}_x}{2}\right) ^ {\otimes N}$, we obtain:
\begin{align}
    \frac{\partial}{\partial \theta_i} \left( \langle \psi (\vec{\theta}) \ket{+} \right)
    & =
    \frac{1}{2} \frac{1}{\langle \psi (\vec{\theta}) \ket{+}} \frac{\partial}{\partial \theta_i} \langle \psi (\vec{\theta}) | \hat{M} |\psi (\vec{\theta}) \rangle\\
    & =
    \frac{1}{2} \frac{1}{\langle \psi (\vec{\theta}) \ket{+}} \frac{\partial}{\partial \theta_i} \expval{\hat{M}}(\vec{\theta})\\
    & = \frac{1}{4}\frac{\expval{\hat{M}}(\theta_i + \frac{\pi}{2}) 
    - 
    \expval{\hat{M}}(\theta_i - \frac{\pi}{2})}{\langle \psi (\vec{\theta}) \ket{+}}
    \label{eq:paramshiftM}
\end{align}
where we have used the parameter shift rule in Eq.~\eqref{eq:paramshiftM} to evaluate $\frac{\partial}{\partial \theta_i} \expval{\hat{M}}$.
Evaluating the gradient $\frac{\partial f_{\textrm{v}}}{\partial \theta_i}$ therefore requires $3$ state preparations per variational parameter, at $\theta_i$, $\theta_i + \frac{\pi}{2}$, and at $\theta_i - \frac{\pi}{2}$.

\section{Quantum Approximate Optimization Algorithm}
\label{appqaoa}

\begin{figure}
\begin{tabular}{cc}
\includegraphics[width=0.45\columnwidth]{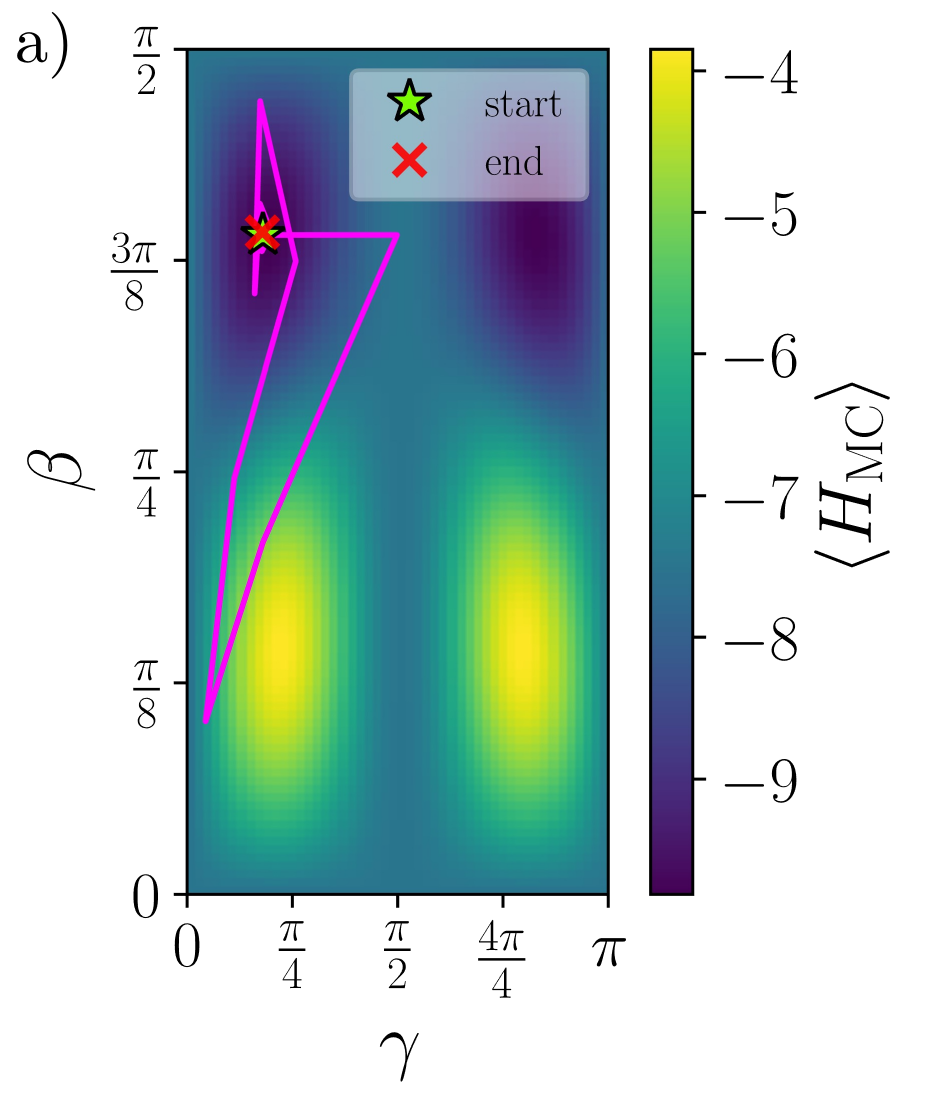} &
\includegraphics[width=0.45\columnwidth]{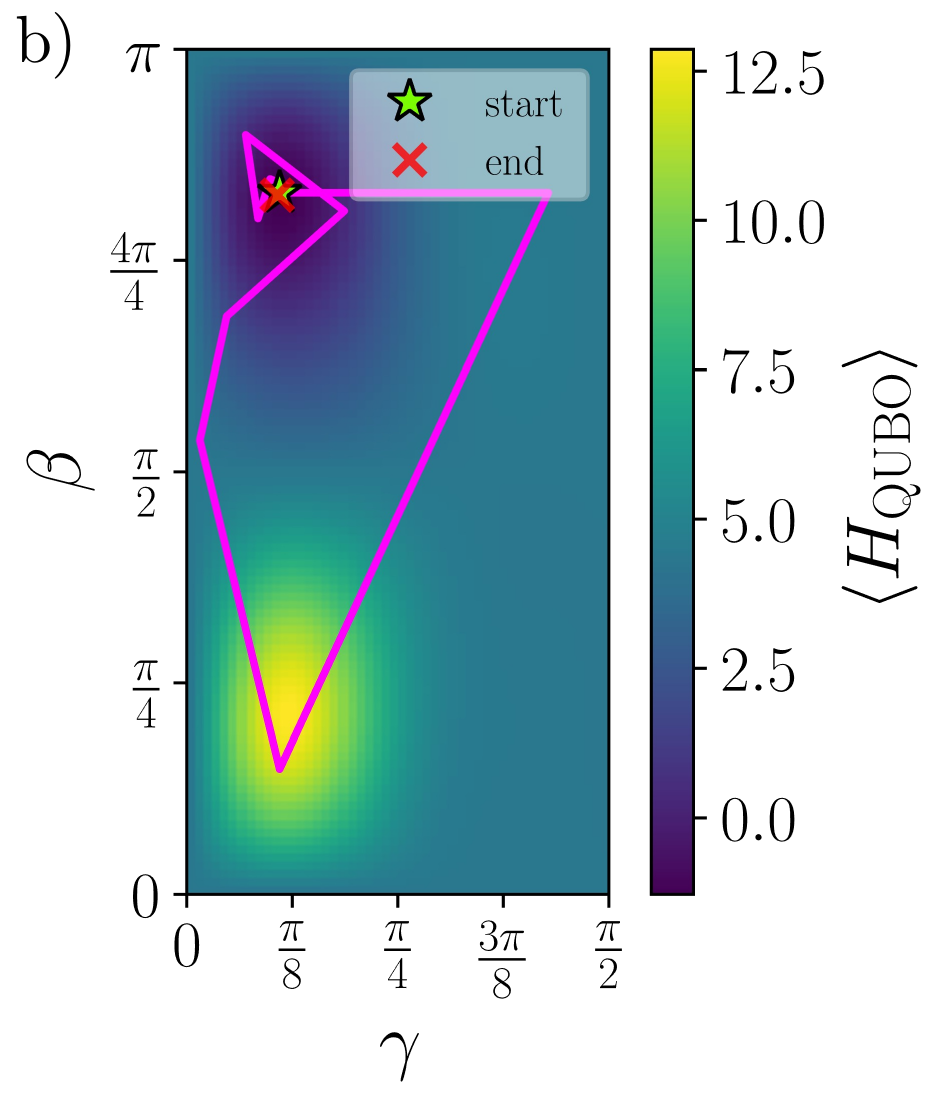}
\end{tabular}
    \caption{Gridsearch for the QAOA with $p=1$ for (a) the MaxCut problem and (b) a randomly generated QUBO problem. The expectation value $\expval{\hat{H}_{\textrm{Ising}}}$ for the respective problems were calculated for a $100 \times 100$ grid using the statevector simulator in Pennylane. The optimal parameters found using the gridsearch (green star) were used as starting parameters for further fine tuning of $\expval{\hat{H}_{\textrm{Ising}}}$ using COBYLA (magenta line). Red cross denotes the final parameters obtained using COBYLA.}
    \label{fig:QaoaGS}
\end{figure}

The quantum approximate optimization algorithm (QAOA) is a variational quantum algorithm for finding approximate solutions to combinatorial optimization problems. 
The QAOA state is parameterized by two sets of angles, $\vec{\gamma} = \{\gamma_1 ,... ,  \gamma_p\}$ and $\vec{\beta} = \{\beta_1 ,... ,  \beta_p\}$:
\begin{equation}
    \ket{\psi(\vec{\gamma}, \vec{\beta})}_{\textrm{QAOA}} = \prod_i^p U_x(\beta_i) U_H(\gamma_i) \ket{+}
    \label{eq:qaoastate}
\end{equation}
where
\begin{align}
    U_H(\gamma) & = e^{-i \gamma \hat{H}_{\textrm{Ising}}} \\
    U_x(\beta) & = e^{-i \beta  \sum_i \hat{\sigma}_i^x}.
\end{align}

In Sec.~\ref{optresults}, we compared the results obtained from preparing the landscape state $\ket{u}$ with results obtained from $p=1$ of QAOA.
For $p=1$, the state in Eq.~\eqref{eq:qaoastate} only contains 2 variational parameters, and the optimal parameters to obtain the QAOA results in Fig~\ref{fig:ResultsLL} were found by using a grid search with a resolution of $100 \times 100$ and then using COBYLA to perform a local search, further fine tuning the parameters. 
Figure~\ref{fig:QaoaGS} shows the grid and fine tuning needed to obtain the optimal parameters.

The variational ansatz described in Sec.~\ref{optresults} to produce the results in Fig.~\ref{fig:ResultsLL} uses $4\times (N-1) = 36$ CNOT gates.
By comparison, decomposing the unitaries in the QAOA to similar gatesets require $2\times n_e$ number of CNOT gates per depth $p$, where $n_e$ is the number of edges in the problem graph \cite{majumdar2021optimizing}.
For $p=1$, this amounts to $30$ and $90$ CNOT gates for the MaxCut and random QUBO graph before accounting for measures to handle long ranged interactions between qubits. 

% \section{Effect of hyperparameters for the MaxCut instance}
% \label{degenhyperparam}

% \begin{figure}
% \includegraphics[width=0.97\columnwidth]{MaxCut_gxplot.pdf} 
%     \caption{Probability of sampling a solution with Hamming distance $d$ away from \textit{any} optimal solution for a $3$-regular MaxCut problem for different values of $\Gamma$ and $\lambda$.}
%     \label{fig:hyperparamhdmaxcut}
% \end{figure}

\end{document}